\documentstyle[]{article}

\input{psfig.sty}
\def\be{\begin{equation}}
\def\ee{\end{equation}}
\def\bea{\begin{eqnarray}}
\def\eea{\end{eqnarray}}

\begin{document}
\vspace{-1.1cm}
\begin{center}
{\Large \bf
Screened Perturbation Theory }\\
\vspace*{0.5cm}
{\large  F. Karsch\\
Fakult\"at f\"ur Physik, Universit\"at Bielefeld,\\
D-33615, Bielefeld,
Germany \vspace{0.5cm}\\
A. Patk\'os and P. Petreczky\\
Department of Atomic Physics, E\"otv\"os University,\\
H-1088, Budapest,
Hungary} \\
\end{center}

\begin{abstract}
A new perturbative scheme is proposed for the evaluation of
the free energy density of field theories at finite temperature.
The screened loop expansion takes into account exactly the phenomenon 
of screening in thermal propagators. The approach is tested 
in the $N$-component scalar field theory at 2-loop level and also at
3-loop in the large $N$ limit. The perturbative series generated
by the screened loop expansion shows much
better numerical convergence than previous expansions
generated in powers of the quartic coupling.  
\end{abstract}

\section{Introduction}
The weak coupling expansion of the QCD free energy density does show 
very bad 
convergence properties. The coefficients of the 
usual perturbative series are of alternating sign and of increasing 
magnitude 
\cite{Shu78,Kap79,Kal81,Toi83,Arn95}. 
Only in the TeV temperature 
range can one find a satisfactory numerical 
convergence rate \cite{Nie96}. This is
in great contrast to numerical calculations of this quantity (or of 
its 
derivative, the internal energy) in Monte Carlo simulations. These
calculations indicate that deviations from the high temperature ideal 
gas 
limit are within 15\% already for temperatures about $5 T_c$ ($T_c$
being
the
critical temperature ) 
\cite{Kar93}.

In the effective theory approach of Braaten and Nieto the contribution
of
different scales, $2 \pi T$, $g T$ and $g^2 T$ to the QCD free energy
was
separated\cite{Bra96}. It was noticed that the apparent bad convergence of 
perturbation theory for the QCD free energy is due to the poor convergence
of the perturbative contribution to the free energy from the momentum
interval $gT<<k<<2 \pi T$ for temperatures few times the
critical \cite{Nie96}. One of the possible reasons for
this breakdown might be the  
fact that scales $2 \pi T$ and $g T$ are not actually separated in the
above
temperature range. This observation lead us to the conjecture that the 
convergence of perturbation series might be improved  if it is
reorganized in a way that does not assume separation of the 
scales $g T$ and $2 \pi T$.

There are several mass scales in QCD, what makes the practical
application of this conjecture more complicated than it looks like at
first sight. On the other hand the bad convergence of the perturbative
free energy density series 
is also seen in the $\phi^4$ scalar field theory and our
aim is to check the above conjecture in this case. In the scalar field
theory the only relevant mass scale is the Debye screening mass .
This scale is taken into account exactly in our approach which we thus
call {\em screened perturbation theory}.

\section{Loop Expansion with Screened Propagators}

We consider an $O(N)$ symmetric scalar field theory with the following
Lagrangian:
\begin{eqnarray}
&
L=L_0+L_{int},\nonumber\\
&
L_0={1\over 2}({(\partial \phi_i)}^2+m^2 \phi^2_i),\nonumber\\
&
L_{int}=-{1\over 2} m^2 \phi_i^2+{g^2\over 24 N} {(\phi_i^2)}^2+
{g^2\over 24 N} (Z_2-1){(\phi_i^2)}^2.
\end{eqnarray}
Following Refs. 9 and 10 we have introduced and subtracted 
a thermal mass term with $m\sim {\cal O} (g)$ which modifies the bare
propagator and thus reorganizes the perturbative series.
The coupling constant renormalisation factor is given by\cite{Amit}
\begin{equation}
Z_2=1+{3 g^2\over {(4 \pi)}^2} {N+8\over 18 \epsilon} + 
{\cal
O}(g^4),
\end{equation}
and we have left out the field renormalisation factor $Z_1$, which is 
unity
up to ${\cal O}(g^4)$.

Our aim is to perform a loop expansion with massive propagators without
making any assumption on the magnitude of $m$, i.e. to evaluate the free
energy and other thermodynamic quantities "exactly" in $m$. This
means that we intend to perform the loop expansion starting from a  
massive ideal gas.

It should be noticed that in the present approach as well as in Ref.9 and
10 static and non-static modes are both resummed. In Ref.12 it was
shown that when the perturbative series is organized in powers of $g$, i.e
when one takes into account that $m \sim {\cal O}(g)$ the resummation
of all modes is equivalent to the resummation of the static mode only.
In our case, however, these two resummation schemes will lead to different
results.

For finite N we have carried out our program up to
2-loop level, for large $N$ it is possible to estimate also the 3-loop
contribution \cite{Karsch97}. The contribution of different diagrams to
the free energy can be found in Ref. 14, where the necessary
technical details are also given. Here we just mention that the contribution
of all diagrams to the free energy are proportional to 1-loop
sum-integrals, except the basketball diagram which is, however, suppressed
in the large $N$ limit. The first problem which arises in the present
approach is that 1-loop integrals entering the free energy contain
divergent and scale dependent terms, with coefficients proportional 
to positive
powers of the screening mass and therefore can not be subtracted or 
canceled by the renormalisation procedure applied at $T=0$. 
There is also a scale dependent
term $\sim g^4 T^4$ in the 3-loop free energy, but this is canceled by
renormalisation of the coupling constant.

The next question to be discussed is the choice of the screening mass.
In the conventional resummation schemes $m$ was chosen to be 
equal to its 1-loop value
\cite{Arn95,Fren92,Par92,Par95} . In our approach the free
energy is evaluated exactly with respect to the screening mass.
Therefore, to be consistent, the screening mass itself should be evaluated
"exactly" in $m$, which means that it should be determined from the
1-loop gap equation, which becomes exact in the large $N$ limit
\cite{Karsch97,Dol74}.  The screening
mass obtained this way agrees well with the 1-loop value for $g\le 1$
while it is only half as large as the 1-loop value for $g \sim 10$.

\begin{figure}
\psfig{figure=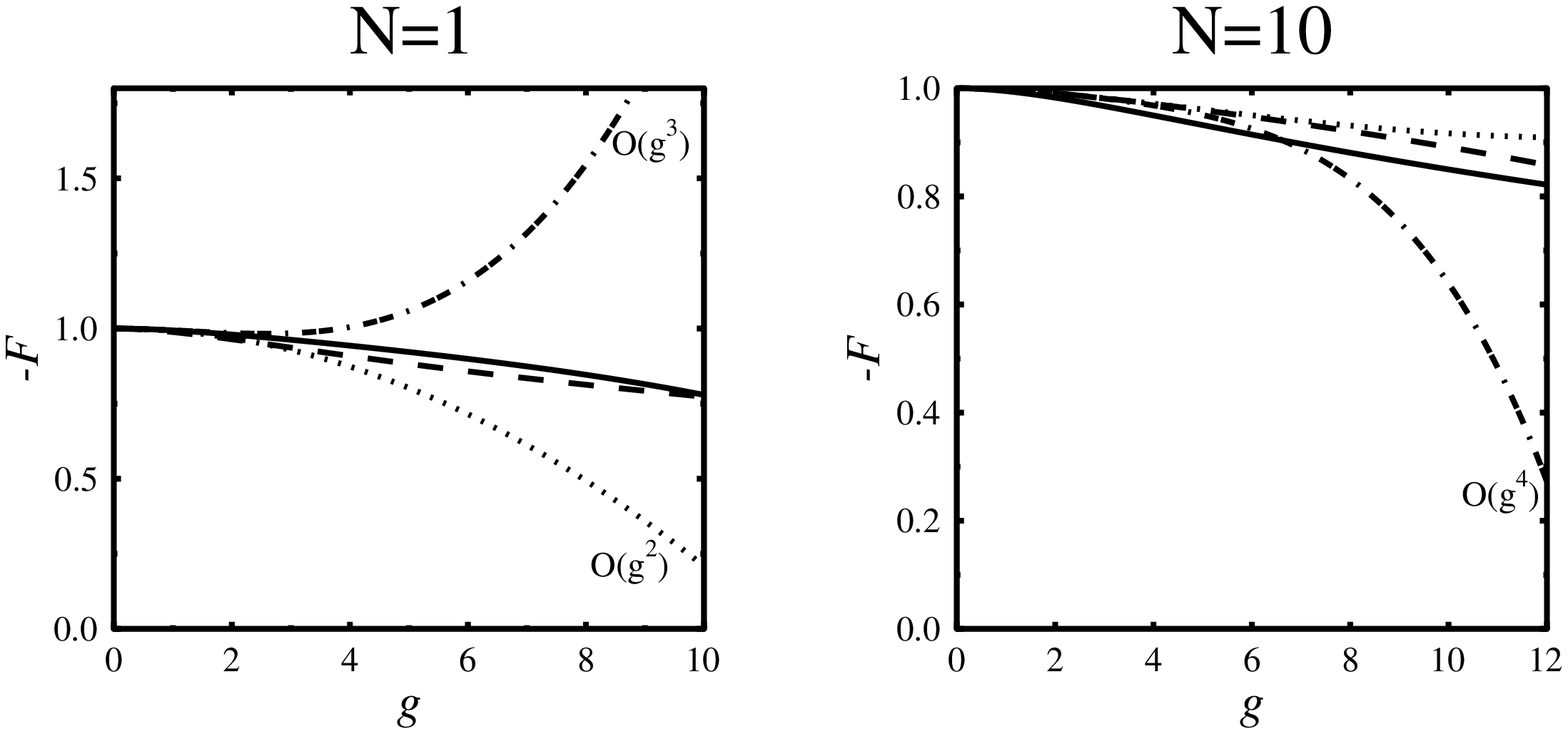,height=5truecm}
\caption{The free energy density of scalar field theory as a function of
the scalar self-coupling g in units of the free energy of a massless ideal
gas ($N {\pi^2 T^4\over 90}$). For N=1 1-loop (dashed line) and 2-loop
(solid line ) results of the screened loop expansion are shown versus the
${\cal O}(g^2)$ and ${\cal O}(g^3)$ results of the conventional approach.
For N=10 the 1-loop (dotted line), 2-loop (dashed line) and 3-loop (solid
line) results are shown versus the ${\cal O}(g^4)$ result of the
conventional approach.
\label{fig:radish}}
\end{figure}

In our  numerical investigation we have studied the cases
$N=1$, and $N=10$
for the large $N$ limit. The free energy for both cases is shown in Fig.1
The screening masses have been computed from the 1-loop gap equation.
For $N=1$ 1-loop and 2-loop results of screened perturbation theory are
shown together with ${\cal O}(g^2)$ and ${\cal O}(g^3)$ result of the
conventional perturbation theory. For $N=10$ 1-loop, 2-loop and 3-loop
results of the screened loop expansion are shown. For comparison we also
show here the conventional ${\cal O}(g^4)$ result. Lower orders ${\cal
O}(g^2)$ and ${\cal O}(g^3)$ do show alternating behaviour similar for the 
$N=1$ case.

\begin{figure}
\psfig{figure=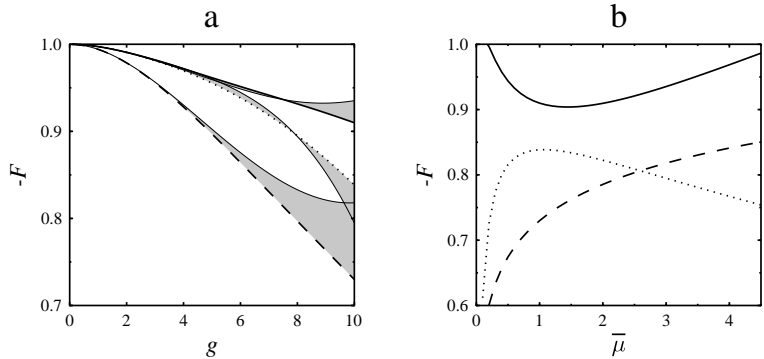,height=5truecm}
\caption{The free energy density of scalar field theory 
in units of the free energy of a massless ideal
gas ($N {\pi^2 T^4\over 90}$).  The 1-loop (dashed line), 2-loop
(dotted line ) 
and 3-loop (solid line) free energy as function of : a) scalar self
coupling $g$ at $\bar \mu =1.5 T$ , b) as function of $\bar \mu$
for $g=10$.
\label{mub}}
\end{figure}

\section{Discussion}
As it was discussed in the previous section at
each order of the screened loop expansion divergent and scale
dependent terms appear, but they are canceled by higher loop
contributions \cite{Karsch97}. However divergent and scale dependent
terms are defined up to a constant, which defines the renormalisation
scheme. In Ref. 14 $\overline {MS}$ was used and terms like $\ln
\bar \mu/T$ were omitted. We can have some insight into the sensitivity of
the free energy on the scheme  keeping the scale dependent terms and
studying the dependence of the final result on the choice of the
renormalisation scale. In Fig.~2a the free energy evaluated in 1-to-3 loop
approximation is shown as function of the scalar
self-coupling for $\bar \mu=1.5$ and $N=10$, the area painted in grey
shows the variation of the free energy as the scale is varied from $\bar
\mu=1.5$  to $\bar \mu=3$. In Fig.~2b the free energy is shown for $g=10$
as a function of $\bar \mu$.
For these figures the leading result for the screening mass was used. 
It should be noticed that scale dependent
part $\sim T^4$ in the 3-loop contribution is canceled by the
renormalisation of the coupling constant. As one can see from the
figures the scale dependence weakens as one goes from the 1-loop to the
3-loop level free energy. The reason for this is that at 3-loop level the 
conjectured cancellation of the scale dependent terms proportional to
$m^{6}$ does occur.

Now let us turn to the question of the choice of the screening mass. One
may think that good convergence of the present approach is due to the
fact that terms like $g^3$ are replaced by $m^3$ and $m$ being
determined from the gap equation is small.  
However, one can see from Fig.~2b that the convergence of screened
perturbation theory is also good 
when the leading order mass is used in the calculation of the free
energy.
The difference between the 3-loop free energy evaluated with
the leading mass and that evaluated with the mass determined
from the 1-loop gap equation is less than $1\%$.

In conclusion, we managed to improve substantially the convergence of
the perturbative series of the $\phi^4$ theory 
using the screened perturbation expansion and we hope
that some of the lessons learned in the scalar field theory remain
valid also in QCD.

\vskip 0.5truecm
\noindent
{\large \bf Acknowledgement}
\vskip 0.3truecm
A.P thanks support form OTKA, grant No. T22929.
P.P thanks J.Pol\'onyi, Z.Fodor and A.Jakov\'ac for useful discussion.



\begin{thebibliography}{17}
\bibitem{Shu78}E. Shuryak, JETP, {\bf 47} (1978) 212
\bibitem{Kap79}J. Kapusta, Nucl. Phys. {\bf B148}, (1979) 461
\bibitem{Kal81}K. Kalashnikov and J. Klimov, Sov. J. Nucl. Phys. {\bf
33} (1981) 647
\bibitem{Toi83}A. Toimela, Phys. Lett. {\bf 124B} (1983) 407
\bibitem{Arn95}P. Arnold and  C. Zhai, Phys. Rev. {\bf D50} (1994)
7603 and 
Phys. Rev. {\bf D51} (1995) 1906,
B. Kastening and C. Zhai, Phys. Rev. {\bf D52} (1995)
7232
\bibitem{Nie96}A. Nieto, Perturbative QCD at High Temperature 
OHSTPY-HEP-T-96-019, hep-ph/9612291, see also the contribution to these
proceedings
\bibitem{Kar93} 
G. Boyd, J. Engels, F. Karsch, E. Laermann, C. Legeland, M. 
L\"utgemeier,
B. Petersson, Nucl.Phys. {\bf B469} (1996) 419
\bibitem{Bra96}
E. Braaten and A. Nieto, Phys. Rev. {\bf D53} (1996)
3421
\bibitem{Fren92}
J. Frenkel, A. Saa, and J. Taylor, Phys. Rev. {\bf D46} (1992) 3670
\bibitem{Par92}
R. Parwani, Phys. Rev. {\bf D45} (1992) 4965
\bibitem{Amit}
D.J. Amit,
Field Theory, Renormalisation Group and \\
Critical Phenomena, chapter 7, p.173
(1978, McGaw-Hill Inc).
\bibitem{Arn93}
P. Arnold and R. Espinosa, Phys. Rev. {\bf D47} (1993) 3546
\bibitem{Par95}
R. Parwani and H. Singh, Phys. Rev. {\bf D51} (1995) 
4518
\bibitem{Karsch97}
F. Karsch, A. Patk\'os, P. Petreczky, Phys.Lett.  {\bf B401} (1997) 69
\bibitem{Schulz97}
J. Reinbach and H.Schulz, hep-ph/9703414
\bibitem{Dol74}
L. Dolan and R. Jackiw, Phys. Rev. {\bf D9} (1974) 3320
\end{thebibliography}
\end{document}